\documentclass[journal]{IEEEtran}
\usepackage{cite}
\usepackage{graphicx}
\usepackage{amsthm,amsmath}
\usepackage{framed}
\usepackage{color}
\usepackage{marvosym}
\theoremstyle{definition} 

\newcommand{\mb}{\,\mathbf}

\newcommand{\eq}[1]{\,\begin{equation}
                   #1 
                   \end{equation}
}

\newcommand{\morabba}[1]{\,\begin{flushright}
 \Rectsteel \\
\end{flushright}}

\newcommand{\eqq}[2]{\,\begin{equation} \label{#2}
                   #1 
                   \end{equation}
}

\newcommand{\mike}[1]{{\color{black} #1}}

\date{}
\title{The Effect of Exogenous Inputs and Defiant Agents on Opinion Dynamics with Local and Global Interactions}

\author{Babak~Fotouhi and Michael~G.~Rabbat\thanks{The authors are with the Department of Electrical and Computer Engineering, McGill University, 3480 University Street, Montr\'eal, Qu\'ebec, Canada, H3A 0E9. Email: babak.fotouhi@mail.mcgill.ca, michael.rabbat@mcgill.ca.}}
\begin{document}

\maketitle
\begin{abstract}
Most of the conventional models for opinion dynamics mainly account for a fully local influence, where myopic agents decide their actions after they interact with other agents that are adjacent to them. For example, in the case of social interactions, this includes family, friends, and other \mike{immediate} strong social ties. The model proposed in this \mike{paper} embodies a global influence as well where by \mike{\emph{global}} we mean that each node also observes a sample of the average behavior of the entire \mike{population; e.g., in the social example, people observe other people on the streets, subway, and other social venues.}  We consider \mike{the} case where nodes have dichotomous \mike{states; examples of applications include elections with two major parties, whether or not to adopt a new technology or product, and any yes/no opinion such as in voting on a referendum.} The dynamics of states on a network with arbitrary degree distribution are studied. For a given initial condition, we find the probability to reach consensus on each state and the expected time reach to consensus. \mike{To model mass media, the effect of an exogenous bias on the average orientation of the system is investigated.} To do so, we add an external field to the model that favors one of the states over the other. This field interferes with the regular decision process of each node and creates a constant probability to lean towards one of the states. We solve for the average state of the system as a function of time for given initial conditions. Then anti-conformists (stubborn nodes who never revise their states) are added to the network, in an effort to circumvent the external bias. We find necessary conditions on the number of these defiant nodes required to cancel the effect of the external bias. Our analysis is based on a mean field approximation of the agent opinions.
\end{abstract}

\section{Introduction}
In many complex network problems originating from the social and economic sciences, the aim is to study and understand the collective network behavior under particular microdynamics. Opinion dynamics models are a subset of these efforts; agent-based models are studied to elucidate the macro-behavior, given a prescribed interaction scheme.
In opinion dynamics models each node is endowed with a state, which adapts and evolves based on observations of the states of other nodes in the network. The state is usually a number, which can be considered continuous~\cite{olfati} or discrete~\cite{krapivsky1}. 
Depending on the problem, nodes may represent, e.g., people~\cite{martins1}, firms~\cite{deroian}, or countries~\cite{goyal}. At each time step, each agent observes the states of other agents with which it interacts, and then the agents update their states based on these observations. 
 
In this paper we study the effects of different interaction phenomenon on the behavior of opinion dynamics models running over networks. Each agent has a binary state $\pm 1$. Examples of cases where such a model may be applicable are: elections with two major parties~\cite{naim, lambiotte}, deciding whether or not to adopt a new technology~\cite{ delre2, delre1}, whether or not to watch a new movie~\cite{delre4}, whether or not to follow a new fashion trend~\cite{redner_fad}, and any other yes/no proposition (go to war or not, referendums, etc.). 

To model the social factors that effect each agent's state, we initially consider two different sources of influence:
\begin{itemize}
\item Local influence: Each agent is influenced by its direct neighbors in the network (e.g., family, close friends and relatives, and other strong social ties).

 \item Global influence:  Each agent is influenced indirectly by the average opinion of the population (e.g., by observing people on the streets, classes, subway, cinemas, and other social places). For example, one takes a sample of the popularity of a fashion trend or  a new technological product (such as smart phones or laptops) by observing other people in the society. The existence of a strong tie is not necessary in this case. Another major example of the global type of influence is the internet.
\end{itemize}
We model both of these sources of influence. At each time step, each node observes the fraction of its neighboring nodes who have each of the two states. These fractions constitute the local part of the social influence, as will be explained in the next section. Each node also observes the fraction of all the nodes  who adopt each of the two states. The global part of the social influence will be based on these fractions.  

We find an expression for the evolution of the average density of each state over time (\mike{i.e., the fraction of nodes who adopt $+1$ and who adopt $-1$}) as a function of the initial conditions. We consider a network with arbitrary degree distribution and conduct our analysis under the mean field approximation. We also find the probability that all nodes will end up having the same state (i.e., reach consensus), and we calculate the expected times to reach consensus on either of the two states. The conventional (or ``pure'') voter model~\cite{liggett1} is a limiting case of our setup, where the global influence is zero, and our results can be seen as generalizing those of~\cite{redner2} for the pure voter model. 

Next, we consider the effect of an exogenous field that influences all agents, biasing them towards $+1$. For example, this could model the effects of mass media. We find an expression for the evolution of the expected state over \mike{time as} a function of the initial conditions. This expression reveals the rate of convergence of all nodes \mike{to the state $+1$ depending on the magnitude} of the external field. 

Finally, we incorporate and study \mike{the effect of anti-conformists (a.k.a stubborn nodes, inflexibles, zealots)} who fix their opinion at $-1$, opposing the external field. In this case, there is never a consensus; instead, the network reaches an equilibrium with a certain fraction of agents taking each side. For this model we determine the fraction of stubborn nodes required to sway the majority opinion to $-1$ for an exogenous field of a particular strength.

\subsection{Related Work}

There is a growing literature on opinion dynamics and related models, and we review the most \mike{relevant} related work here. Different opinion dynamics models have different interaction processes and different updating schemes, which are contingent on the specificities of the problem. 
For example, one can choose random blocks of adjacent nodes who enforce their opinion on neighbors under certain conditions \cite{sznajd1,sznajd2,crokidakis1}. In the so-called \emph{voter model},  \mike{at each timestep 
a randomly chosen node copies} the state of a randomly chosen neighbor~\cite{krapivsky1, liggett1, bernardes, lambiotte, volovik, redner1, redner2}.  Conversely, in the \emph{invasion process}, the randomly chosen node imposes its state on a randomly chosen neighbor~\cite{castel1}.  

Learning and trust have been incorporated in the models, where each node has an estimate of how reliable its observations might be~\cite{ozdaglar, acem2, goyal2, martins1,klimek, martins2}. In some models, nodes only interact with neighbors whose opinions are close to their own~\cite{naim1, kraus1, deffuant, weisbuch}. In~\cite{stark}, agents have inertia in the sense that the longer they have had a particular state, the less probable it \mike{is} for them to deviate from it. The reader is also referred to~\cite{castel_review} for a broad review.

In~\cite{mobilia, rednerZ,ozdaglarZ,democracy}, the effect of stubborn  agents (a.k.a., zealots \mike{or inflexibles}) \mike{is} studied. These agents commit to a certain state and never vary. In~\cite{mobilia} the effect of a single zealot is studied, and it is shown that in one and two dimensions the system reaches consensus over the state enforced by the zealot. \mike{However, this is not} true in higher dimensions. In~\cite{rednerZ} the voter model with \mike{an} arbitrary number of zealots is examined \mike{on the complete graph and lattice graphs in one and two dimensions}, and it is shown that the magnetization has \mike{a} Gaussian distribution whose characteristics only depend on the number of zealots \mike{and} not on the entire population. The upshot is that a \mike{small} number of zealots can prevent consensus, regardless of the population size. In~\cite{ozdaglarZ}, the expected values of the average state for arbitrary number of zealots on an arbitrary graph is found for the pure voter model, and it is shown that consensus is prevented as long as stubborn nodes exist. Also, the steady state of the system is shown to be independent of the initial states of the non-stubborn nodes. The interested reader is also referred to~\cite{democracy}, for further simulations and discussions, and also~\cite{galam1,galam2,inflex1, inflex2}, and the broad overview of the models presented in~\cite{galam3}.

In~\cite{redner2} the pure voter model on heterogeneous graphs is solved under the mean-field assumption. \mike{Nodes with identical degrees are considered to be indistinguishable} in dynamics, and the connection probability of each pair of nodes is proportional to their degrees. \mike{The probability to reach consensus on each of the states and the expected time to reach consensus are approximated.}
In~\cite{masuda}, intrinsic ``flip" rates are heterogeneous, so that some nodes adopt new opinions more frequently than others.

\subsection{Paper Organization}

The rest of this paper is organized as follows. The basic model, incorporating both local and global influence, is formally defined and studied in Section~\ref{sec:localGlobal}. Then, Section~\ref{sec:exogenous} studies the effect of an exogenous field. Section~\ref{sec:anticonformists} introduces anti-conformists to the model. Finally, we conclude in Section~\ref{sec:conclusion}. Throughout the paper, theoretical predictions are accompanied by \mike{numerical simulations}.

\section{Dynamics in the Absence of Exogenous Bias} \label{sec:localGlobal}

\subsection{Network Model and Basic Dynamics}

We consider a network of $N$ nodes\footnote{Throughout, we use the terms "agent" and "node" interchangeably.}. The network structure is defined through the neighborhoods of each node. For node $x$, let $N_x$ denote the neighbors of node $x$ (i.e., the set of nodes adjacent to $x$), and let $z_x=|N_x|$ be the degree of node $x$ (the number of its neighbors).  The underlying network is assumed to be undirected: for two nodes $x$ and $y$, we have $y \in N_x$ if and only if $x \in N_y$. The network is also assumed to be connected: for any two agents $x$ and $y$, there is a sequence of nodes $x_0 = x, x_1, \dots, x_\ell = y$ such that $x_k \in N_{x_{k-1}}$ for each $k=1, \dots, \ell$. We will be particularly interested in random network \mike{models such as} Erdos-Renyi graphs~\cite{ER1,ER2}, preferential attachment graphs~\cite{barabasi}, and random recursive trees~(as in~\cite{rednerG, RRT}, with $A_k=1$).

Each node has a binary-valued state $s_x(t) \in \{-1, +1\}$. Nodes have initial opinions $s_x(0)$ at time $t=0$, and time progresses forwards. At each time step, one node is picked randomly, and it updates its state. In general, the update will involve the node either keeping the same state, or flipping to the other state. 
In order to avoid burdensome notation below, we will omit the dependence on time when it is clear from the context. Let us denote the state of the system by a vector $\vec{\mathbf{s}}=(s_1, s_2, \dots, s_N)$. 
The probability that a node $x$ flips its state at time $t$ is given by
\[
P\{s_x(t + \delta t) = - s_x(t) | s_x(t)\} = w_x(\vec{\mathbf{s}}(t)),
\]
where 
\eq{
w_x(\vec{\mb{s}})=\frac{\theta}{2} \left( 1-\frac{s_x}{z_x} \sum_{y\in N_x} s_y \right) +\frac{1-\theta}{2}
 \left( 1-\frac{s_x}{N} \sum_{\textnormal{all }y} s_y \right)
.}
The first term, with weight $\theta \in [0,1]$, captures the local influence. It is equal to the fraction of neighbors who oppose node $x$.  The second part is the global influence, with weight $1-\theta$. It equals the fraction of the entire population that disagree with node $x$. Here $\theta$ is a tradeoff parameter between the influence of close social ties and the influence of the society. Node $x$, upon being selected, flips its state with probability $w_x$. Hence, if all nodes disagree with $x$ then $x$ flips its state with probability 1. On the other hand, if all nodes agree with $x$ then $x$ does not flip its state with probability $1$.

\subsection{A Conservation Law and Consensus Probabilities}

Now we focus on finding the probability that nodes reach  consensus \mike{on $+1$ and on $-1$}. Note that when the system is not unanimous, at each \mike{time step} there are nonzero probabilities to move towards the states $+1$ or $-1$ since, as long as consensus has not been reached, there is at least one node with nonzero flipping probability which, by construction, has nonzero probability for being picked to update its state. Hence the only absorbing states of the system are those where either all nodes have $+1$ or $-1$. 

We solve the model in continuous-time approximation, examining the system as $\delta t \rightarrow 0$, in which case for any quantity $f(t)$ we have $\frac{f(t + \delta t) - f(t)}{\delta t} \rightarrow \dot{f} = \frac{\partial f}{\partial t}$. Let $c_x$ denote the expected value of $s_x$. The evolution of $c_x$ is given by
\eqq{
\dot{c}_x =-c_x+\frac{\theta}{z_x}\sum_{y\in N_x} c_y + \left( \frac{1-\theta}{N} \sum_{\textnormal{all }y} c_y \right)
.}{c1}

Let $n_k$ denote the fraction of nodes with degree $k$, and let $\bar{z} = \frac{1}{N} \sum_x z_x = \sum_k k n_k$ denote the average degree of the network. Also define $\rho_k(t)$ to be the fraction of nodes with degree $k$ which have $s_x(t) = +1$. Consider the quantity $\mu(t) \stackrel{\text{def}}{=} \sum_k k n_k \rho_k$, which can be interpreted as the expected value of the quantity $k \rho_k$ for a graph with degree distribution $\{n_k\}$. 

Under the assumption that the degrees of adjacent nodes are uncorrelated, the probability that two nodes are connected is proportional to the product of their degrees; i.e.,
\begin{equation}
P\{y \in N_x | z_y = k\} = \frac{k z_x}{N \bar{z}}, \label{prop}
\end{equation}
and $P\{y \in N_x\} = z_y / N$. Also recall that $n_k = P\{z_y = k\}$. Then by Bayes' rule, we have
\[
P\{z_y = k | y \in N_x\} = \frac{\frac{k z_x}{N \bar{z}} \cdot n_k}{\frac{z_x}{N}} = \frac{k n_k}{\bar{z}}.
\]
The assumption that the degrees of adjacent nodes are uncorrelated is justified in a number of scenarios. For example, this is the case in the classical Erdos-Renyi random graph model~\cite{ER1, ER2}. Social networks have also been observed to have preferential attachment properties, and in the well-known scale-free network model of Barabasi and Albert~\cite{barabasi} edges are no longer independent since new edges connect to high-degree nodes with higher probability. Nonetheless, in Section~\ref{subsec:BA} we justify that \eqref{prop} still holds true for scale-free graphs.

We will also make a mean-field assumption and approximate that, for a neighbor $y \in N_x$ with degree $z_y = k$, the expected state $c_y$ is equal to $2\rho_k - 1$, the average state of all nodes with degree $k$. Under these approximations, we have
\begin{align}
\sum_{y \in N_x} c_y &\approx z_x \sum_k (2\rho_k - 1) \cdot \frac{k n_k}{\bar{z}} \nonumber \\
&= z_x \left(\sum_k \frac{2 k \rho_k n_k}{\bar{z}} - \sum_k \frac{k n_k}{\bar{z}}\right) \nonumber \\
&= z_x \left(\frac{2 \mu}{\bar{z}} - 1\right). \label{sum_neighbors}
\end{align}

Let $m(t) = \frac{1}{N} \sum_x s_x(t)$ denote the average state at time $t$. Then combining (\ref{c1}) and \eqref{sum_neighbors} gives
the following differential equation for $m(t)$:
\eqq{
\dot{m}(t)=-m(t)+\theta \left( \frac{2\mu}{\bar{z}}-1 \right) + (1-\theta) m(t)
.}{mdot}

Multiplying \eqref{c1} by $z_x$ and summing over all nodes gives the following expression for the evolution of $\mu(t)$:
\eqq{
\dot{\mu}(t)= \left[ \frac{\bar{z}}{2}(1-\theta) \right] m(t) - (1-\theta) \mu(t) + \left(\frac{\bar{z}}{2}(1-\theta)\right)
.}{mudot}

Comparing (\ref{mdot}) with (\ref{mudot}), observe that
\eqq{
\dot{m}(t) \left[ \frac{\bar{z}}{2} (1-\theta) \right] + \theta \dot{\mu}(t)=0
.}{mt1}
Thus, we arrive at the following conservation law: 
\eqq{
\psi(t) \stackrel{\text{def}}{=} m(t)\left[ \frac{\bar{z}}{2} (1-\theta) \right] + \theta \mu(t)= \textnormal{const.}
}{psi1}
For the pure voter model (i.e., with only local influences), we have $\theta=1$ and $\psi=\mu$ is conserved, which matches the findings of~\cite{redner1}.

Now this conservation law can be used to directly determine the probability that the network reaches consensus on a particular state, given the initial opinions at each node. Note that $\psi$ equals $\frac{\bar{z}}{2} (1+\theta)$ for the case where $s_x=+1$ for all $x$, and $\psi = \frac{- \bar{z}}{2}(1 - \theta)$ when $s_x = -1$ for all $x$. 
For brevity let us define:
\eq{
\begin{cases}
P^u \stackrel{\text{def}}{=} P\{m(\infty)=1\}  \\
P^d \stackrel{\text{def}}{=}  P\{m(\infty)=-1\} \ .
\end{cases}
}
Using the fact that 
\[
\psi(0) = \lim_{t \rightarrow \infty} \psi(t) = P^u \cdot \frac{\bar{z}}{2} (1+\theta) + (1 - P^u) \cdot \frac{- \bar{z}}{2}(1 - \theta),
\]
the probability of reaching consensus on \mike{$+1$} is easily found to be 
\eqq{
\displaystyle
\mike{P^u =
 \frac{1-\theta}{2}+\frac{\psi(0)}{\bar{z}}}
.}{Pu1}
Similarly, we find \mike{that the probability of reaching consensus on $-1$ is}
\eqq{
P^d =
\frac{1+\theta}{2}-\frac{\psi(0)}{\bar{z}}
.}{Pd1}

In terms of the initial states, we can rewrite \eqref{Pu1} as
\eqq{
P^u=\frac{1}{2} + \frac{1-\theta}{2N}\sum_x s_x(0) + 
\frac{\theta}{2N\bar{z}} \sum_x z_x s_x(0) 
,}{PU0}
and \eqref{Pd1} also can be rewritten as
\eqq{
P^d=\frac{1}{2}- \frac{1-\theta}{2N}\sum_x s_x(0)
 -\frac{\theta}{2N\bar{z}} \sum_x z_x s_x(0) 
.}{PD0}

\mike{Note that the obtained exit probability (i.e., the probability of reaching consensus on $+1$) is a continuous function of the initial fraction of nodes with $+1$ as $N \rightarrow \infty$. This is in   contrast to other models for which the exit probability tends to a step-like function; see~\cite{exit_p} and references therein.}

\begin{figure}[ht]
  \centering
  \includegraphics[width=90mm, height=45mm]{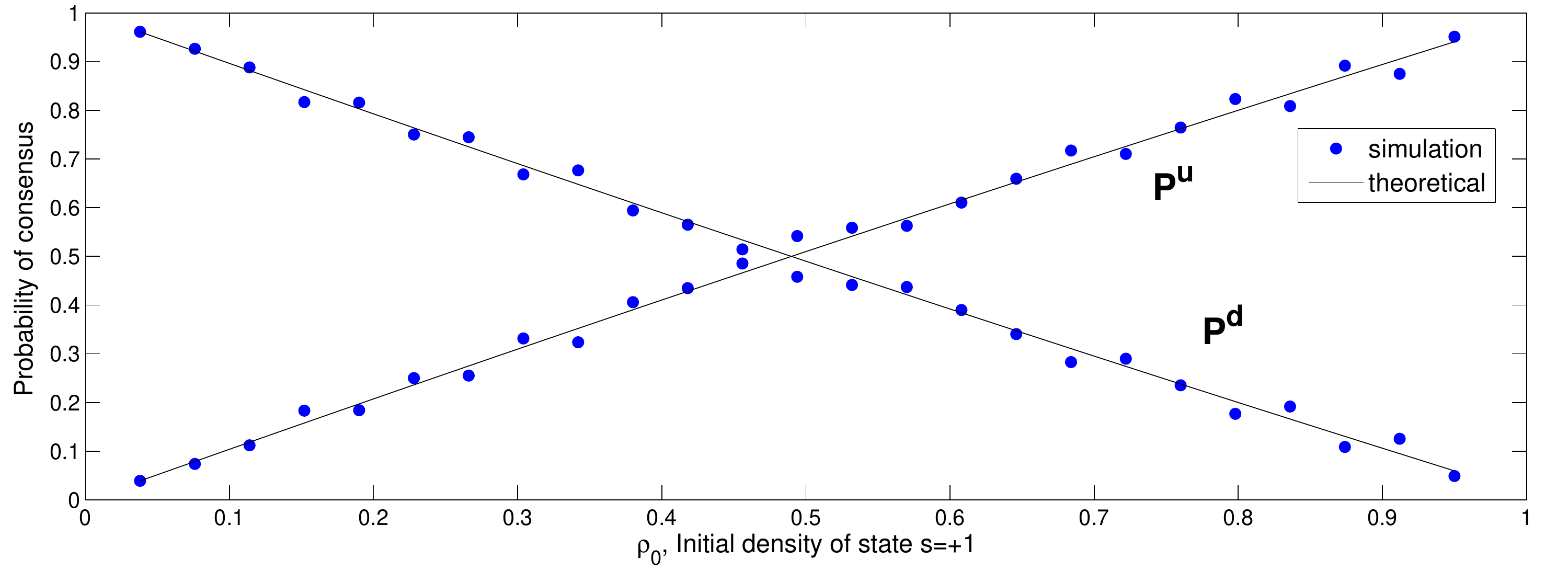}
  \caption[Figure 1]%
  {The expected probability of reaching consensus on the state $+1$ for different initial conditions for the fraction of nodes with $s=+1$, on a network 
of 400 nodes,  for $\theta=0.7$. The results are averaged over 1000 Monte Carlo simulations. The underlying graph is a Barabasi-Albert scale-free graph~\cite{barabasi} with $m=2$. Theoretical prediction is given by~\eqref{PU0} and~\eqref{PD0}.}
\label{figP}
\end{figure}

Figure~\ref{figP} illustrates the probabilities $P^u$ and $P^d$ for reaching consensus on $\pm 1$ as a function of initial fraction \mike{of nodes with $+1$} for opinion dynamics run on Barabasi-Albert scale free graphs with $400$ nodes. The initial opinions at each node are drawn i.i.d., with the initial density of state $+1$ varying between $0.05$ and $0.95$, and the local-global tradeoff parameter \mike{$\theta = 0.7$}. As expected, the probability $P^u$ to reach consensus on $+1$ is higher when more nodes initially have state $+1$, and this probability is linear in the initial density of nodes with $+1$.

Next, we can solve for the dynamics of the average state, $m(t)$. Equations (\ref{mdot}) and (\ref{mudot}) are a standard system of linear differential equations (see~\cite{boyce,simmons, braun}).  The solution for $\mu(t)$ is
\eqq{
\mu(t)=\left[ \psi(0) +\frac{\bar{z}}{2} (1-\theta) \right] + \left[ \mu(0) - \psi(0) - \frac{\bar{z}}{2}(1-\theta) \right] e^{-t}
,}{mu_plot_1}
and for $m(t)$ we have
\eqq{
m(t)=\left[ \frac{2}{\bar{z}}\psi(0) -\theta \right] + \left[ m(0) - \frac{2}{\bar{z}}\psi(0) + \theta \right] e^{-t}
.}{m_plot_1}

\begin{figure}[ht]
  \centering
  \includegraphics[width=90mm, height=45mm]{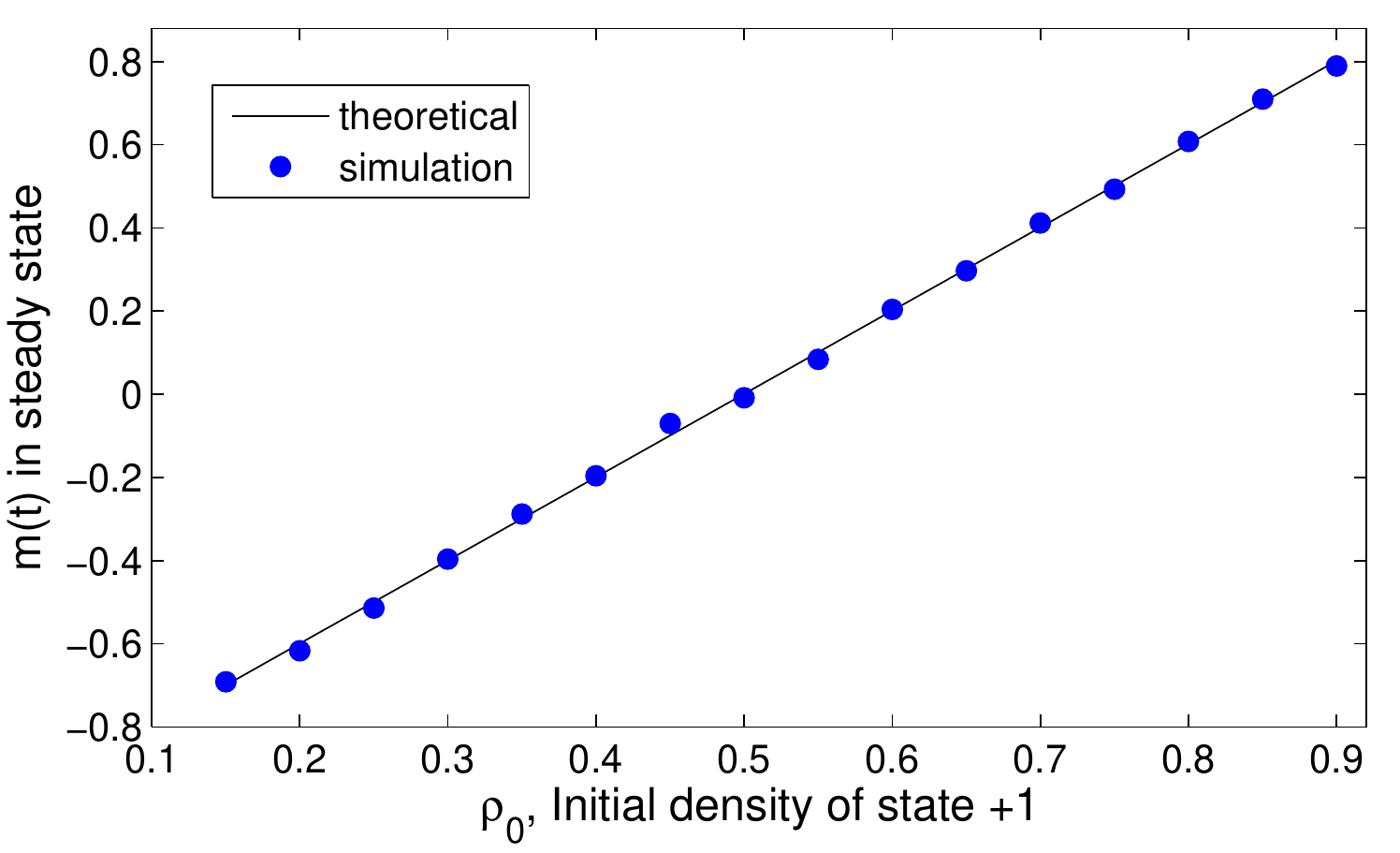}
  \caption[Figure 2]%
  {Steady state values of $m(t)$ as a function as $\rho(0)$ for different values of $\rho(0)$ between 
$0.1,0.9$ (uniform) and the predicted value of equation (\ref{mt1}) with $\theta=0.7$. The underlying graph is a random recursive tree (as in~\cite{rednerG}, with $A_k=1$) with 200 nodes. Results are averaged over 100 Monte Carlo trials. The theoretical prediction is given by~\eqref{m_plot_1} in the limit of~${t\rightarrow \infty}$. }
\label{figM}
\end{figure}

\begin{figure}[ht]
  \centering
  \includegraphics[width=90mm, height=45mm]{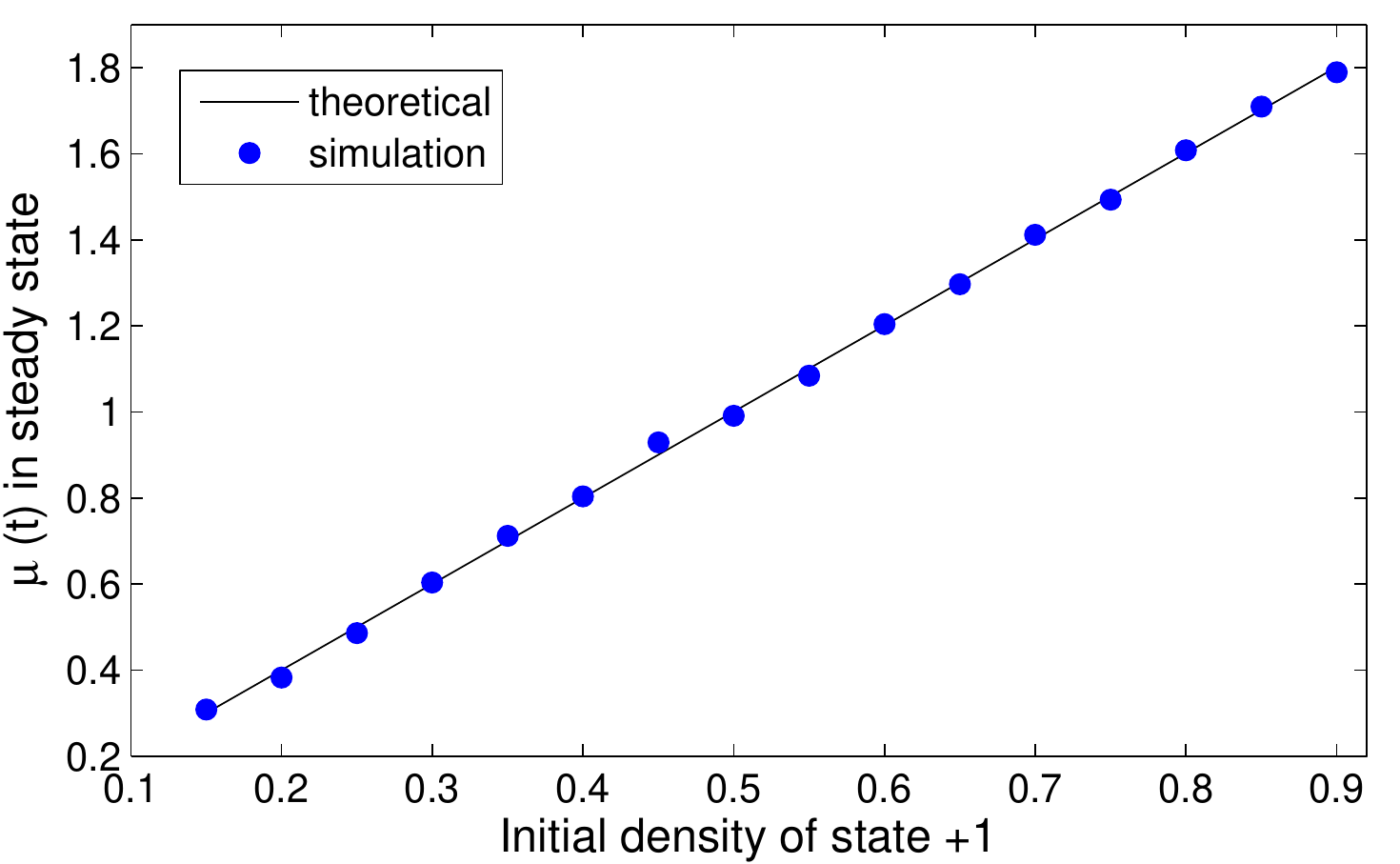}
  \caption[Figure 2]%
  {Steady state values of $\mu(t)$ as a function as $\rho(0)$ for different values of $\rho(0)$ between 
$0.1,0.9$ (uniform) and the predicted value of equation (\ref{mt1}) with $\theta=0.7$. The underlying graph is a random recursive tree (as in~\cite{rednerG}, with $A_k=1$) with 200 nodes. Results are averaged over 100 Monte Carlo trials. The theoretical prediction is given by~\eqref{mu_plot_1} in the limit of~${t\rightarrow \infty}$.}
\label{figMU}
\end{figure}

For both $m(t)$ and $\mu(t)$ there is a steady-state value and there is a transient exponentially-decaying part. Figures~\ref{figM} and~\ref{figMU} are depictions of the theoretical prediction and simulation results for the steady-state values.

Note that for $\theta=0$ (i.e., purely global influence, which is equivalent to a complete graph), equation (\ref{psi1}) implies that $m$ is conserved. Also for $\theta=1$ the conserved quantity is $\mu$. Both of these polar cases agree with the results \mike{reported in~\cite{redner2}. Moreover, the expressions derived here allow us to interpolate between these two extremes}.

\subsection{Time to Consensus}

Now let us focus on the expected time \mike{to reach consensus}. Recall that $n_k$ is the fraction of nodes with degree $k$; thus $N n_k$ is the number of nodes with degree $k$. Of the nodes with degree $k$, if one changes state from $-1$ to $+1$, then note that $\rho_k$ changes by $1 / (N n_k)$. Let us define $\delta \rho_k = \frac{1}{N n_k}$, and define
\eq{
\begin{cases}
P^+_k  \stackrel{\text{def}}{=} P\{ \rho_k \rightarrow \rho_k + \delta \rho_k \}  \\
P^-_k \stackrel{\text{def}}{=}   P\{ \rho_k \rightarrow \rho_k - \delta \rho_k \}  
\end{cases}
.}
For the first one we have
\begin{align}
P^+_k &=n_k (1-\rho_k) \left[ \frac{1}{2}+ \frac{\theta}{2}\left( \frac{2\mu}{\bar{z}}-1\right) + \frac{1-\theta}{2}m \right] \nonumber \\
&=n_k (1-\rho_k) \left(\frac{1-\theta}{2}+\frac{\theta}{\bar{z}}
\mu + \frac{1-\theta}{2} m \right)
,
\end{align}
which implies that a degree-\mike{$k$} node with state $-1$ is \mike{picked and} it flips its state at this step. Similarly, 
\begin{align}
P^-_k &=n_k  \rho_k \left[ \frac{1}{2}- \frac{\theta}{2}\left( \frac{2\mu}{\bar{z}}-1\right) - \frac{1-\theta}{2}m \right] \nonumber \\
&=n_k \rho_k \left(\frac{1+\theta}{2}-\frac{\theta}{\bar{z}}
\mu - \frac{1-\theta}{2} m \right)
.
\end{align}

Let us denote by $T(\vec{\rho})$ the expected consensus time when the densities for population of nodes with various degrees are $\vec{\rho} = (\rho_1, \rho_2, \rho_3, \ldots)$. Note that $T(\vec{\rho})$ is the expected time to reach consensus on either state. We will find the expected time to reach consensus on individual states $+1$ or $-1$ afterwards. 
The expected consensus time satisfies the following recurrence relation: 
\begin{align}
T(\vec{\rho}) = \delta t &+ \sum_k \left[ T(\vec{\rho}+\delta \rho_k \hat{k}) P^+_k +  T(\vec{\rho}-\delta \rho_k \hat{k}) P^-_k \right] \nonumber \\
&+ \left[ 1- \sum_k  (P^+_k +  P^-_k) \right] T(\vec{\rho})
.
\label{trho1}
\end{align}
The first term is the time increment due to an update occurring. The second term accounts for the expected time to consensus, given that the selected node has changed its state. The variable $k$ in the sum runs over all possible degrees in the network. The last term is the expected time to consensus, given that the selected node has made no change in its state.

Let us normalize time units so that $\delta t= \frac{1}{N}$. 
Define the constants
\eqq{
\begin{cases}
 \lambda \stackrel{\text{def}}{=} \frac{1-\theta}{2}+\frac{\psi}{\bar{z}}\\
\alpha \stackrel{\text{def}}{=} \theta - \frac{2\psi}{\bar{z}}
\end{cases}
.}{l_a}
Since $\psi$ is conserved, we will drop the argument $t$. 

\begin{figure}[ht]
  \centering
  \includegraphics[width=90mm, height=45mm]{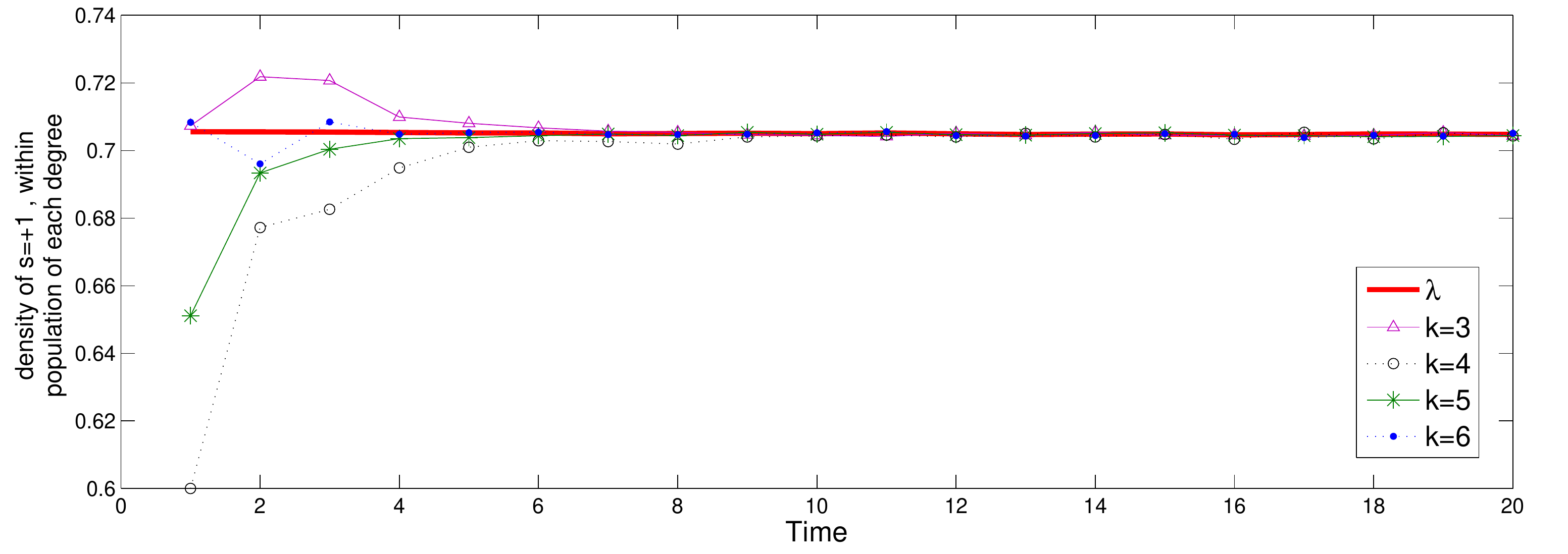}
  \caption[Figure 2]%
  {Simulation results compared to what~\ref{rhobardot1} predicts. Results are averaged over 1500 Monte Carlo simulations (with identical initial conditions), for a Barabasi-Albert graph~\cite{barabasi} with $N=500$ nodes and $m=2$. The initial condition was $\rho=0.7$, with nodes selected randomly regardless of degrees. The value of $\theta$ is $0.7$. The theoretical prediction is extracted from~\eqref{rhobardot1}.}
\label{rhos_lambda}
\end{figure}

Let us denote ${\frac{\partial}{\partial \rho_k} T(\vec{\rho})}$ by $\partial_k T(\vec{\rho})$ for brevity.  In the continuous approximation (as $N \rightarrow \infty$), equation~\eqref{trho1} becomes
\eqq{
\sum_k (\lambda - \rho_k) \partial_k T(\vec{\rho}) + \sum_k \frac{\delta \rho_k}{2}\left[ \lambda + \alpha \rho_k \right]  \partial^2_k T(\vec{\rho}) = -1
.}{T1}
\mike{To simplify this equation we note that the change in $\bar{\rho}_k$, the expected value of $\rho_k$. is given by}
\begin{align}
\frac{d}{dt} \bar{\rho}_k &= P^+_k-P^-_k= n_k (\lambda-\bar{\rho}_k) \nonumber \\
& \Longrightarrow \bar{\rho}_k(t)=\lambda + (\rho_k(0) - \lambda) e^{-n_k t}
,
\label{rhobardot1}
\end{align}
which \mike{implies} that the deviation of $\rho_k$ from the central value $\lambda$ decays exponentially, and thus the expected density $\bar{\rho}_k$ rapidly approaches $\lambda$ for all $k$. It is observed in the simulations that this phase is very short compared to the rest of the process (see Table \ref{table_transient}).  Also Figure~\ref{rhos_lambda} shows samples of the time behavior of $\rho_k$ in simulation.

\begin{table}[t]
\centering 
\begin{tabular}{| c |c |c |c|} 
\hline 
 N & $T$ & $T_{\lambda}$ & $\frac{T_{\lambda}}{T}$ \\ [0.5ex] 
\hline 
 100 &  61  & 3 & 0.049\\ 
 200 & 149 & 3 &  0.020\\
 300 & 221 & 5 &  0.023\\
 400 & 317 & 6 &  0.019\\
 500 & 402 & 8 &  0.020\\ 
 600 & 445 & 8 &  0.018\\ 
 700 & 496 & 9 &  0.018\\ 
 800 & 549 & 12 & 0.022\\ 
 900 & 638 & 14 & 0.022\\ 
 1000& 755 & 15 & 0.020 \\[1ex] 
\hline 
\end{tabular}
\caption{Comparing the average time to reach consensus (denoted by $T$), and the short initial phase required for nodes of each degree to reach  $\rho_k=\lambda(1 \pm 0.03 )$ (denoted by $T_{\lambda}$). The value of $\theta$ is chosen randomly in the interval of $(0.1,0.4)$ for each simulation. $T_{\lambda}$ is averaged over all degrees. The underlying graph is a scale-free Barabasi-Albert graph as in \cite{barabasi}, with $m=2$. The quantities are averaged over $100$ Monte Carlo simulations for each case. The initial distribution of states is generated randomly for each simulation.} 
\label{table_transient} 
\end{table}

Using~\eqref{rhobardot1}, the first term in~\eqref{T1} vanishes, and~\eqref{T1} simplifies to:
\eqq{
\sum_k \frac{\delta \rho_k}{2}\left[ \lambda + \alpha \rho_k \right]  \partial^2_k T(\vec{\rho}) = -1
, }{T2}
 or equivalently, 
\begin{align}
&\sum_k \left( \frac{1-\theta}{2}+\frac{\psi}{\bar{z}} \right)  \left( \frac{1+\theta}{2}-\frac{\psi}{\bar{z}} \right) \nonumber \\
& \times \frac{n_k}{N}
 \left[ \theta k + \bar{z} (1-\theta) \right]^2 \partial^2_\psi T = -1
.
\end{align}


Note that when every node has $s=+1$ then the time to consensus is zero. In this case we have $m=1$ and $\mu=\bar{z}$. Similarly, the time to consensus is also zero when all nodes have $s=-1$, in which case $m=-1$ and $\mu=0$.  These observations give us the boundary conditions required to solve the differential equation:
\eqq{
\begin{cases}
T\left( \psi= \frac{\bar{z}}{2}(1+\theta) \right) = 0 \\
T\left( \psi= \frac{-\bar{z}}{2}(1-\theta) \right) = 0
\end{cases}
.}{Tbound}

Let us define the constant
\eqq{
A\stackrel{\text{def}}{=} -\frac{N}{\theta^2 \frac{\langle z^2\rangle}{\bar{z}^2}+1-\theta^2}
,}{A_def}
where $\langle z^2\rangle = \frac{1}{N} \sum_x z_x^2$ is the average over all nodes of the degree squared. Then equation \eqref{T2} can be written as
\eqq{
A \left( \frac{1-\theta}{2}+\frac{\psi}{\bar{z}} \right)  \left( \frac{1+\theta}{2}-\frac{\psi}{\bar{z}} \right) \partial^2_\psi T = 1
.}{T3}

The expected consensus time is given by the solution of (\ref{T2}) with the boundary conditions (\ref{Tbound}). The solution is readily obtained by integrating twice:
\begin{align}
T =A & \bigg\{ \left( \frac{1-\theta}{2} + \frac{\psi}{\bar{z}}  \right)  \left[ \ln \left( \frac{1-\theta}{2} + \frac{\psi}{\bar{z}} \right) \right]
\nonumber \\
& + 
\left( \frac{1+\theta}{2} - \frac{\psi}{\bar{z}}  \right)  \left[ \ln \left( \frac{1+\theta}{2} - \frac{\psi}{\bar{z}} \right) \right]
 \bigg\}
.
\end{align}

We can use $P^u$ and $P^d$ to get a shorter expression for $T$:
\eq{
T=A [P^u \ln P^u + P^d \ln P^d] 
,}
or equivalently, 
\eqq{
T=A \bigg[P^u \ln P^u + (1-P^u) \ln (1-P^u)\bigg] 
.}{T_plot_1}
Hence, the time to consensus depends on an entropy-like quantity involving the probability of reaching consensus on either state. Note that $A < 0$. Thus, this expression confirms with the intuition that the expected time to reach consensus is largest when the probability of reaching consensus on either of the states $\pm 1$ is equal, given the initial conditions.

\begin{figure}[ht]
  \centering
  \includegraphics[width=90mm, height=45mm]{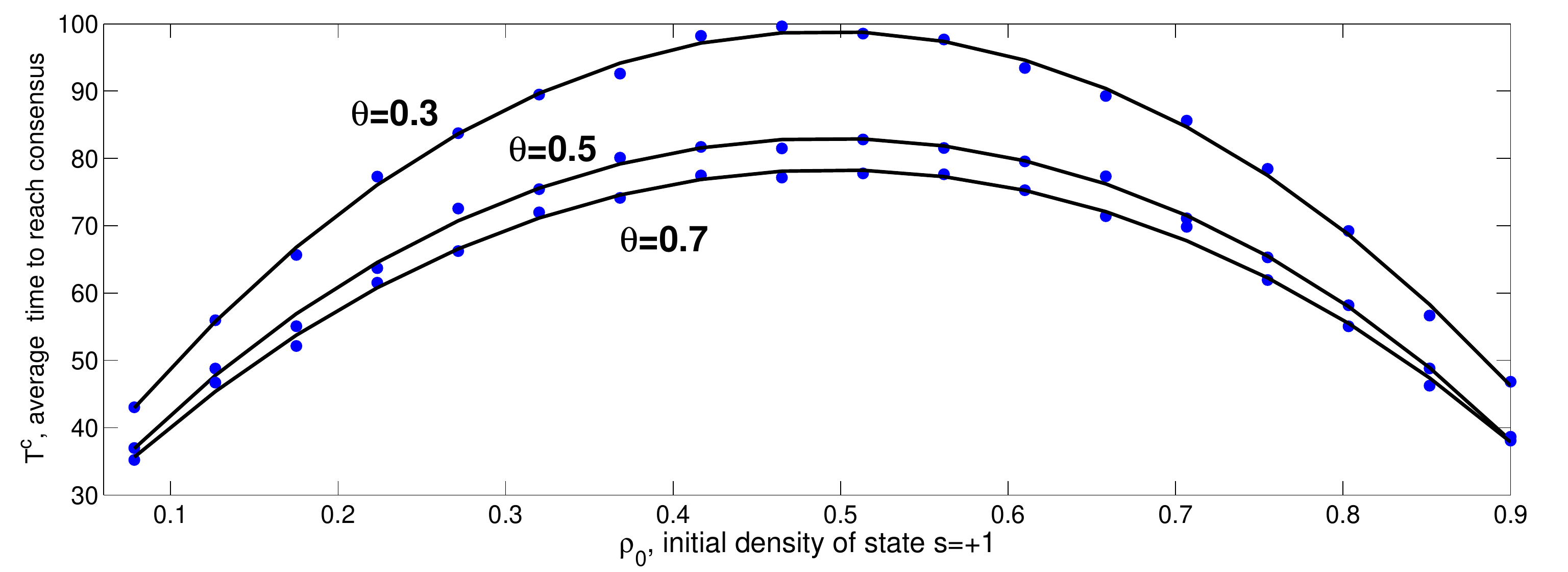}
  \caption[Figure 2]%
  {The expected time to reach consensus for different initial conditions for the fraction of nodes with $s=+1$, on a network 
of 100 nodes, for a Barabasi-Albert graph~\cite{barabasi} with $m=2$, for different values of $\theta$. The simulation results are compared to the prediction of~\eqref{T_plot_1}.}
\label{figT}
\end{figure}

\begin{figure}[ht]
  \centering
  \includegraphics[width=90mm, height=45mm]{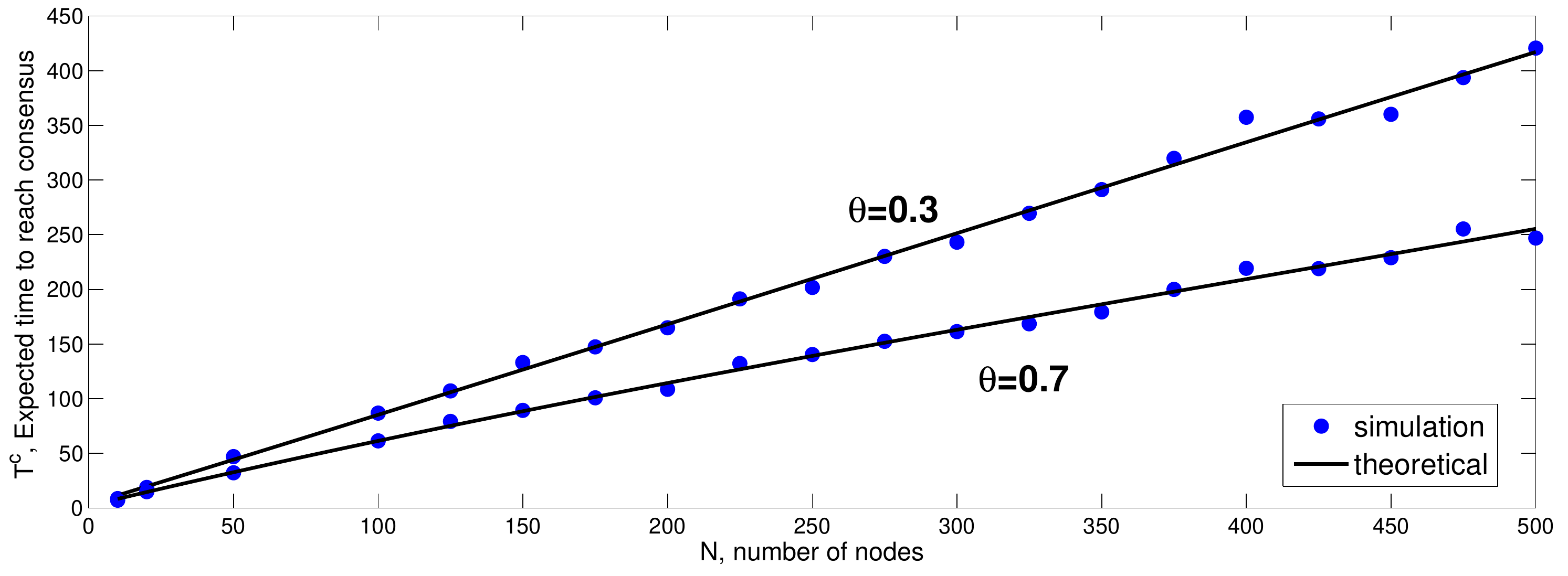}
  \caption[Figure 2]%
  {Consensus time as a function of $N$ for a Barabasi-Albert graph~\cite{barabasi} with $m=2$. The   results are averaged over 500 Monte Carlo simulations. The simulation results are compared to the prediction of~\eqref{T_plot_1}.}
\label{figT_Np}
\end{figure}

The theoretical prediction for $T$ is validated in simulation in Figure \ref{figT}. \mike{Also, Figure~\ref{figT_Np} depicts the consensus time as a function of the population size for a Barabasi-Albert network and random recursive tree (as defined in~\cite{rednerG}, with $A_k=1$).}

Now let us examine two polar cases, $\theta=1$ and $\theta=0$. These polar cases were previously studied in~\cite{redner1} and~\cite{redner2}. When $\theta=1$, we get the conventional voter model as discussed in~\cite{redner2}. From~\eqref{PU0} and~\eqref{A_def} we have 
\eq{
\begin{cases}
\displaystyle A=\frac{-N}{\frac{\langle z^2\rangle}{\bar{z}^2}} 
= -N \frac{\bar{z}^2}{\langle z^2\rangle} \medskip \\ 
\displaystyle P^u= \frac{\mu}{\bar{z}},
\end{cases}
}
which gives 
\eq{
T=-N \frac{\bar{z}^2}{\langle z^2\rangle} 
\bigg[ 
\left( 1-\frac{\mu}{\bar{z}} \right) \ln \left(1- \frac{\mu}{\bar{z}} \right)
+
\left( \frac{\mu}{\bar{z}}\right) \ln \left(\frac{\mu}{\bar{z}} \right) 
\bigg] 
.}
This corresponds to Equation (13) of~\cite{redner2}.

For the other case, $\theta = 0$, we have: 
\eq{
\begin{cases}
\displaystyle A= -N \\
\displaystyle P^u=\rho,
\end{cases}
}
which gives
 \eq{
T=-N 
\bigg[ 
( 1-\rho ) \ln(1- \rho)
+
\rho \ln \rho 
\bigg] 
.}
This matches Equation (25) in~\cite{redner1}. Thus, the expressions derived in this \mike{paper} can be viewed as generalizing the previous work~\cite{redner2,redner1} to interpolate between the cases where agents experience either purely local or purely global influence.

\mike{\subsection{Time to Consensus Conditional on the Final State}}

The analysis above \mike{provides} an expression for $T$, the expected time to reach consensus on any state. Next, we refine our analysis by characterizing the time to consensus given that consensus is reached on a particular state.
Denote by $T^u$ the time to reach consensus given that consensus is reached on $+1$, and let $T^d$ be defined analogously for consensus on $-1$. Similar to~(\ref{trho1}), the recurrence relation for $T^u$ becomes:
\begin{align}
P^u(\vec{\rho}) T^u(\vec{\rho}) = &P^u(\vec{\rho}) \delta t + \left[ 1- \sum_k  (P^+_k +  P^-_k) \right] P^u(\vec{\rho}) T^u(\vec{\rho}) \nonumber \\
&+ \sum_k \bigg[ P^u(\vec{\rho}+\delta \rho_k \hat{k}) T^u(\vec{\rho}+\delta \rho_k \hat{k}) P^+_k \nonumber \\
&+ 
P^u(\vec{\rho}-\delta \rho_k \hat{k}) T^u(\vec{\rho}-\delta \rho_k \hat{k}) P^-_k \bigg].
\label{Tu1}
\end{align}
Following a similar line of reasoning that led to~\eqref{T2}, we arrive at the following differential equation: 
\begin{align}
\sum_k & \bigg\{\left( \frac{1-\theta}{2}+\frac{\psi}{\bar{z}} \right) \left( \frac{1+\theta}{2}-\frac{\psi}{\bar{z}} \right) 
\nonumber \\
& \times \frac{n_k}{N}
 \left[ \theta k + \frac{\bar{z}}{2} (1-\theta) \right]^2 \partial^2_\psi P^u T^u \bigg\}= -P^u
.
\label{Tudiff1}
\end{align}
Using~\eqref{A_def}, this further simplifies to:
\eqq{
A \left( \frac{1-\theta}{2}+\frac{\psi}{\bar{z}} \right)  \left( \frac{1+\theta}{2}-\frac{\psi}{\bar{z}} \right) \partial^2_\psi \bigg[P^u T^u \bigg] = P^u
}{Tudiff3}

Now, note that the boundary conditions are different from~\eqref{Tbound}. In the present case, if all nodes have $s=-1$ then the expected time to reach consensus on $s=+1$ goes to infinity. So we have:
\eqq{
\begin{cases}
T^u\left( \psi= \frac{\bar{z}}{2}(1+\theta) \right) = 0 \\
T^u\left( \psi= \frac{\bar{z}}{2}(1-\theta) \right) \rightarrow \infty
\end{cases}
.}{Tubound}
Using these boundary conditions, the solution to~\eqref{Tudiff3} is obtained by integrating twice:
\eqq{
\displaystyle
T^u=A
\frac{\frac{1+\theta}{2} - \frac{\psi}{\bar{z}}  }
{\frac{1-\theta}{2} + \frac{\psi}{\bar{z}}  }
   \ln \left( \frac{1+\theta}{2} - \frac{\psi}{\bar{z}} \right)  
.}{Tu_final}

Similarly, for the expected time to reach consensus given that consensus is reached on $s=-1$, we get: 
\eqq{
\displaystyle
T^d=A
\frac{\frac{1-\theta}{2} + \frac{\psi}{\bar{z}}  }
{\frac{1+\theta}{2} - \frac{\psi}{\bar{z}}  }
   \ln \left( \frac{1-\theta}{2} + \frac{\psi}{\bar{z}} \right)  
.}{Td_final}

\begin{figure}[ht]
  \centering
  \includegraphics[width=90mm, height=45mm]{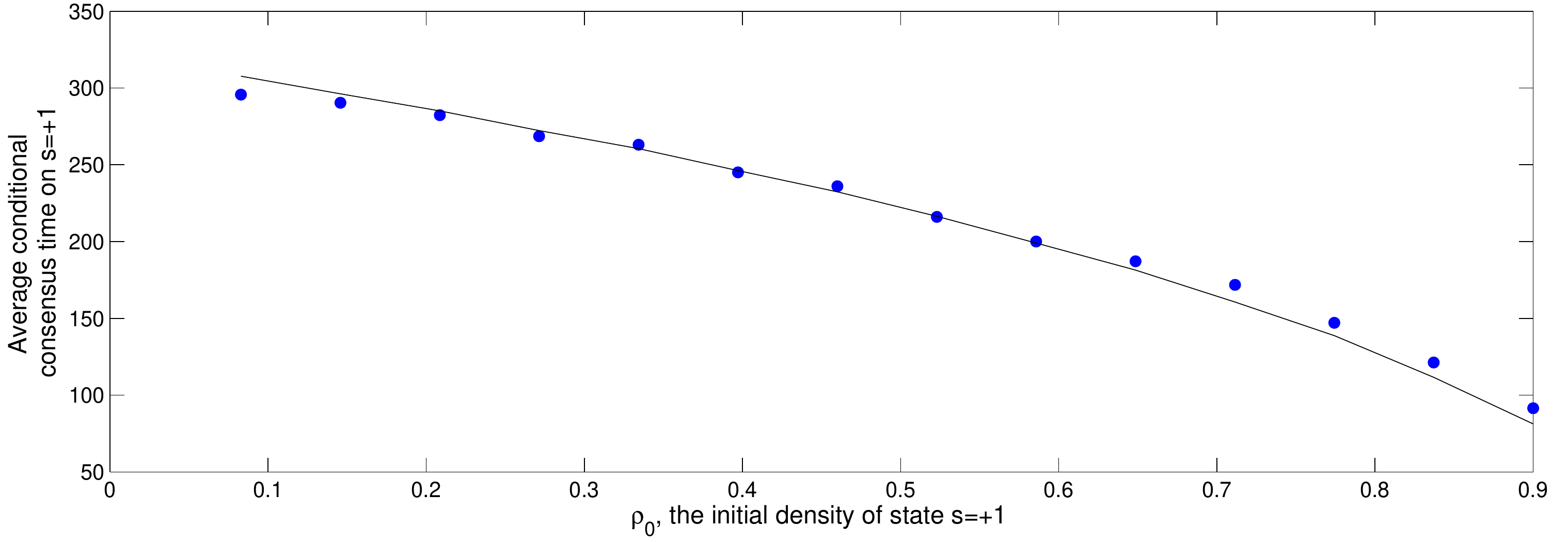}
  \caption[Figure 2]%
  {Consensus time conditional upon consensus over $s=+1$. The underlying graph is a Barabasi-Albert~\cite{barabasi} with $m=2$ with 200 nodes. Results are averaged over 800 simulations. Theoretical prediction is given by~\eqref{Tu_final}.}
\label{Tu_fig}
\end{figure}

\begin{figure}[ht]
  \centering
  \includegraphics[width=90mm, height=45mm]{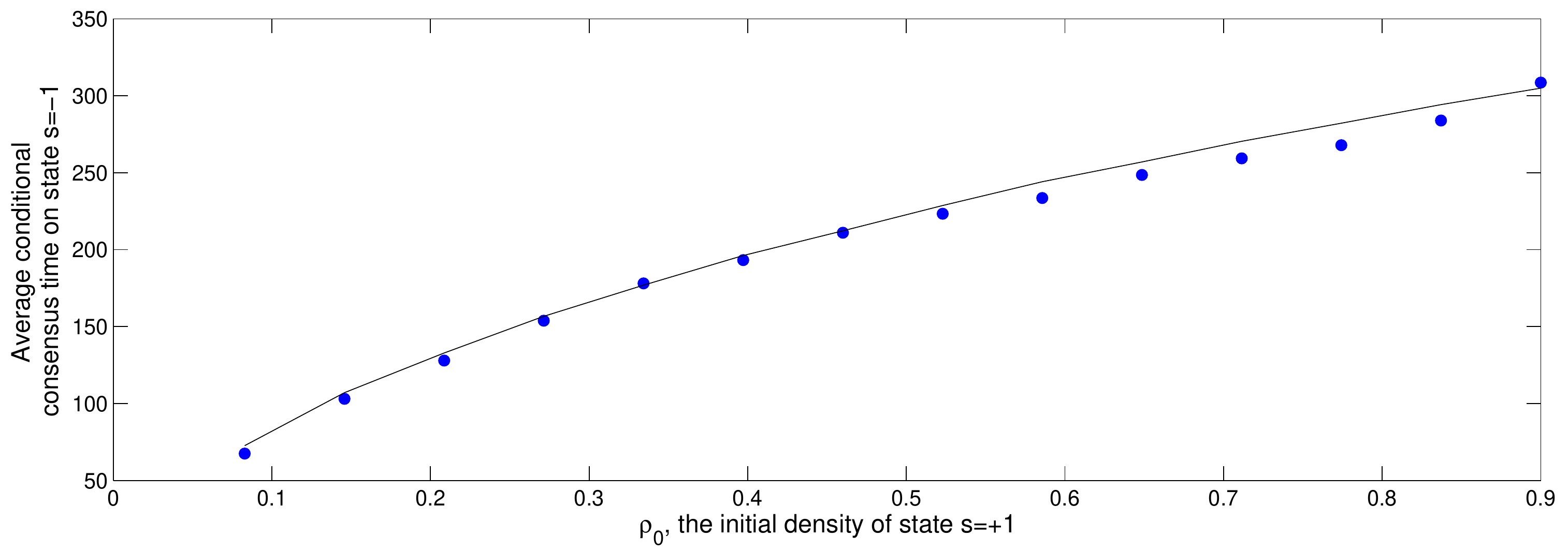}
  \caption[Figure 2]%
  {Consensus time conditional upon consensus over $s=-1$. The underlying graph is a Barabasi-Albert~\cite{barabasi} with $m=2$ with 200 nodes. Results are averaged over 800 simulations. Theoretical prediction is given by~\eqref{Td_final}.}
\label{Td_fig}
\end{figure}

Figures~\ref{Tu_fig} and~\ref{Td_fig} compare the theoretical predictions of \eqref{Tu_final} and \eqref{Td_final} to simulations. 

\mike{One can verify, using \eqref{Pu1}, \eqref{Pd1}, \eqref{Tu_final}, and \eqref{Td_final}, that the following relation holds:}
\eq{
P^u T^u+P^d T^d=T
.}
This implies that $T$ is the expected value of consensus, on a probability space of two events, consensus on $s=+1$ and consensus on $s=-1$ with respective probabilities $P^u$ and $P^d$. Hence, one can also use this relation to extract $T^d$, once $T^u$ and $T$ are obtained, rather than solving differential equation congruent to~\eqref{Tudiff3} for $T^d$, which is obtained by replacing all $P^u$s by $P^d$s, and $T^u$ by $T^d$.  

\subsection{Dynamics on the Barabasi-Albert Graph} \label{subsec:BA}

\mike{Let us examine, in passing,} the validity of (\ref{sum_neighbors}) for the Barabasi-Albert model~\cite{barabasi}. The Barabasi-Albert model is defined through a sequential \mike{algorithm} for constructing a graph. In each step of the algorithm, one new node is introduced and is connected to $m$ existing nodes. (Note that these steps are different than time steps in the opinion dynamics models studied above.) Let us say that node $j$ is introduced at step $j$. Then the probability that node $j$ forms a link with an existing node $i < j$ is
\eqq{ 
P(i \leftrightarrow j)= m \frac{z_i(j)}{\sum_{j' < j} z_{j'}(j)},
}{pref1}
where $z_i(j)$ is the degree of node $i$ just before node $j$ is added to the network.

This scheme is called \emph{preferential attachment} to evoke the notion that existing nodes with higher degrees are more likely to receive new links. The denominator in~(\ref{pref1}) is equal to $j \bar{z}(j) = 2 m j$ (i.e., twice the total number of links in the network when $j$ is added), so we have
\eq{
\frac{d}{dj} z_i(j) =\frac{z_i(j)}{2j} \quad \Longrightarrow \quad z_i(j)=m\sqrt{\frac{j}{i}}
.}

The probability of $j$ linking to $i$  becomes
\eq{
P_t(i \leftrightarrow j)=  \frac{z_i(t) z_j(t)}{2mt} \quad \Longrightarrow\quad P(i \leftrightarrow j)= \frac{z_i z_j}{N \bar{z}}
.}

This result is equivalent to saying that the $ij$-th element of the adjacency matrix is 1 with probability $\frac{z_i z_j}{N \bar{z}}$ and  is zero otherwise.  Using this result, we have
\begin{align}
\sum_{y \in N_x} c_y &= \sum_{\textnormal{all }y} \frac{z_x z_y }{N \bar{z}} c_y \nonumber \\
&=  \sum_k \frac{z_x k}{N \bar{z}} N n_k (2\rho_k -1) \nonumber \\
&= \frac{z_x}{\bar{z}} \left[2\sum_k  k n_k \rho_k  -\sum_k k n_k \right] 
\nonumber \\
&= z_x \left[ \frac{2 \mu}{\bar{z}} -1 \right]
,
\end{align}
which is consistent with (\ref{sum_neighbors}).

\section{Dynamics in the Presence of Exogenous Influence} \label{sec:exogenous}

Now we extend our model by adding an external bias that tries to align all the states towards +1. This external field models the effect of mass media. Examples include advertisements on TV, the internet,  billboards,  theatres, etc. 

Each node assigns a weight to personal observations (which themselves 
are comprised of local and global interactions) and a weight to the external bias. This means that there is a constant probability that each node flips  towards the external field. We denote this weight by $\gamma\in[0,1]$, and the strength of the external field is denoted by $B\in (0,1]$. A larger value of $\gamma$ corresponds to less exposure to the media bias. 
It is clear that eventual consensus on $+1$ is inevitable since $B > 0$. Now we focus on how the dynamics of the system change when the field is introduced. 

We begin by reexamining the conservation law derived in the previous section. The evolution of the expected state \mike{of node $x$} is 
\eqq{
\dot{c}_x=\gamma \left[ -c_x + \frac{\theta}{z_x} \sum_{y\in N_x} s_y + (1-\theta)m \right] + (1-\gamma) B
.}{c_B}

Note that this model is keeping the probability that a node chooses~$s=-1$ intact, while it adds a constant value to the probability that it chooses~$s=+1$. This model is valid while the latter does not exceed unity. The smaller the value of $B$ is, the wider is the range of values of~$m$ for which the model works. For example, if all nodes have~$s=+1$, then the external bias is irrelevant, since the probability of adopting~$s=+1$ is already unity, and cannot exceed any further. 

\mike{Summing~\eqref{c_B} over all nodes} gives
\eqq{
\dot{m}=(-\theta \gamma) m + \left( \frac{ 2 \theta \gamma}{\bar{z}} \right) \mu - \theta \gamma + B(1-\gamma)
.}{m_B}
Multiplying (\ref{c_B}) by $z_x$ and summing over all nodes yields
\eqq{
\dot{\mu}= \frac{\bar{z}\gamma}{2} (1-\theta) m + (1-\theta) \gamma \mu + \frac{\bar{z}}{2} \left[ (1-\theta) \gamma + (1-\gamma) B \right] 
.}{mu_B}

From (\ref{m_B}) and (\ref{mu_B}) we see that $\psi(t)$ is no longer a conserved quantity; rather, it evolves as follows
\eq{
\dot{\psi}=\frac{d}{dt} \left[ \frac{\bar{z}}{2}(1-\theta) m + \theta \mu \right]= \frac{\bar{z}}{2}(1-\gamma)B
,}
 so we have
\eqq{
\psi(t)=\psi(0)+(1-\gamma)\frac{\bar{z}}{2}Bt
. }{psi_with_B}

Using this, we can express $\mu$ in~\eqref{m_B} in terms of $m$ and time, which yields a first order linear equation for $m(t)$. We get:
\eq{
\dot{m}=-\gamma m + \frac{2 \gamma}{\bar{z}}\psi(0) + \gamma (1-\gamma) B t
+ B (1-\gamma)
.}
 Integrating this equation gives the average state over all nodes:
\eqq{
m(t)=B(1-\gamma) t + \left[2\frac{\psi(0)}{\bar{z}}-\theta \right] + \left[ m(0) - 2\frac{\psi(0)}{\bar{z}} + \theta \right] e^{-\gamma t}
.}{m_with_B}
Note that  this expression only holds for values of~$t$ for which~$m(t)<1$; \mike{once $m(t)$ reaches unity, consensus is achieved at $+1$ which is an absorbing state of the system}. For example, to get an estimate of this time (a.k.a the fixation time) for small~$\gamma$, up to the first order we have:
\eq{
m(t)=m(0)+ \left\{B(1-\gamma)-\gamma \left[ m(0) - 2\frac{\psi(0)}{\bar{z}} + \theta \right] \right\} t
,}
so the time for which the average state reaches unity is 
\begin{align}
&t(m=1)= \frac{1-m(0)}{B} \nonumber \\
&+\frac{1-m(0)}{B^2}  \left\{B + m(0) - 2\frac{\psi(0)}{\bar{z}} + \theta  \right\} \gamma + O(\gamma^2)
.
\end{align}
So~$m(t)$ can be expressed in the following compact form: 
\eq{
\begin{cases}
m(t)=\min\{+1,m_1(t)\} \\
m_1(t)\stackrel{\text{def}}{=}B(1-\gamma) t + \left[2\frac{\psi(0)}{\bar{z}}-\theta \right] 
(1-e^{-\gamma t})+  m(0) e^{-\gamma t}
\end{cases}
.}
For $\mu(t)$ we plug \eqref{m_with_B} into \eqref{psi_with_B} to get:
\begin{align}
\mu(t) &=\frac{\gamma \bar{z}}{2} B (1-\gamma) t + \left[ \psi(0)+\frac{\bar{z}}{2}(1-\theta) \right] \nonumber \\
& + \left[ \mu(0) - \psi(0)- \frac{\bar{z}}{2}(1-\theta) \right] e^{-\gamma t}
.
\end{align}
Again, this holds as long as~$\mu(t)\leq \bar{z}$, after which it stays~$\bar{z}$ because it will be an absorbing state of the system.

\mike{In the presence of the external bias, equation (\ref{rhobardot1}) becomes}
\begin{align}
\frac{d}{dt} \bar{\rho}_k &= P^+_k-P^-_k \nonumber \\
&= n_k \bigg\{ \underbrace{\gamma \left( \frac{1-\theta}{2} \right) +B\left( \frac{1-\gamma}{2} \right)  
+ \frac{\gamma \psi(0)}{\bar{z}}}_{\stackrel{\text{def}}{=}   a} 
\nonumber \\
&-  \gamma  \rho_k
+  \underbrace{\frac{\gamma}{2}(1-\gamma)B}_{\stackrel{\text{def}}{=}   b}t\bigg\} 
.
\label{rhodot_B}
\end{align}
Note that \eqref{rhodot_B} is a standard linear differential equation with integration factor $\psi(t)=\exp (n_k \gamma t)$ (for example, see~\cite{arfken, boas}). \mike{Its solution is given by}
\begin{align}
\bar{\rho}_k(t) &= \left( \frac{b}{\gamma} \right) t + \left( \frac{a}{\gamma}-\frac{b}{n_k \gamma^2} \right) \nonumber \\
& + \left( \rho_k(0)
 + \frac{b}{n_k \gamma^2}-\frac{a}{\gamma} \right) e^{-n_k \gamma t}
,
\end{align}
 so the expected densities grow linearly with time and the population is destined to conform to the external bias. 
 
\begin{figure}[ht]
  \centering
  \includegraphics[width=90mm, height=45mm]{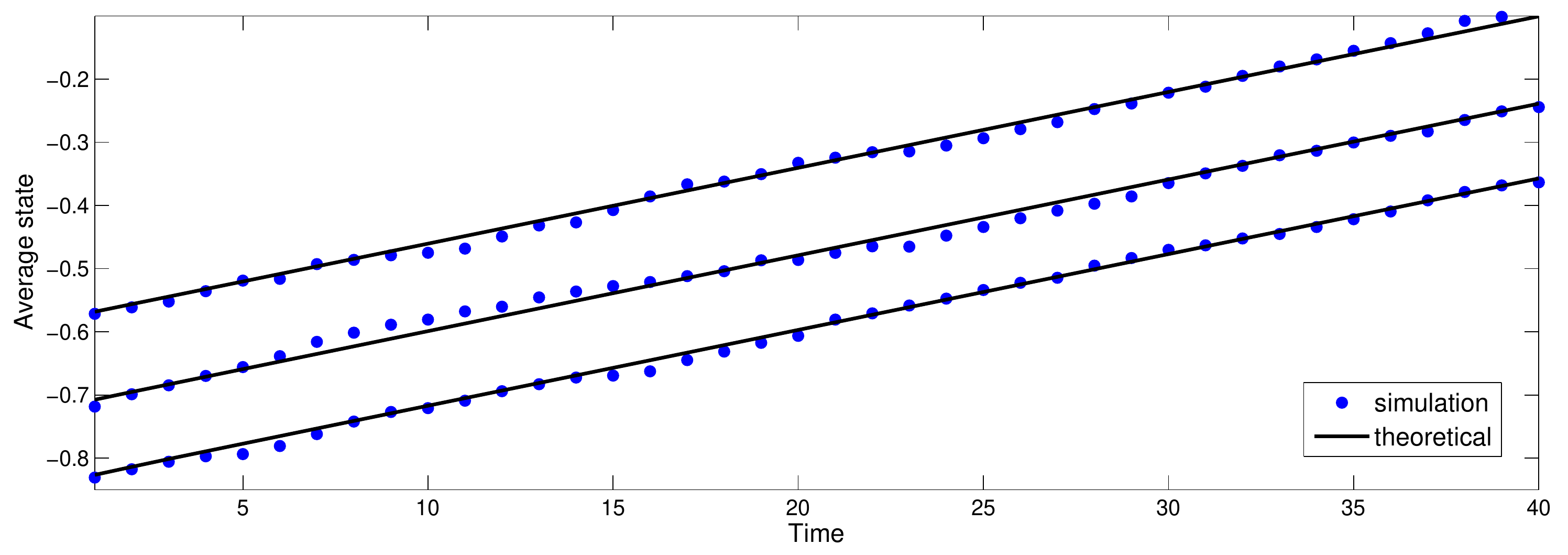}
  \caption[Figure 2]%
  {m(t) with respect to time, for $\rho(0)=.08$ (bottom),$0.14$ (middle) and $0.21$ (top) for $\gamma=0.7$ and $B=0.04$. The underlying graph is a Barabsi-Albert graph~\cite{barabasi} with 200 nodes. Results are averaged over 500 Monte Carlo trials. Theoretical prediction is given by~\eqref{m_with_B}.}
\label{figmb}
\end{figure}

 Figure \ref{figmb} compares the theoretical prediction with simulations. Note that expected states cannot exceed $+1$, so after a finite time we have $s_x=+1$ for all $x$, which means that the whole population complies to the state imposed externally. 

\section{The Effect of Anticonformists} \label{sec:anticonformists}

Now we seek the minimum number of nodes necessary to constantly adopt $s=-1$ in order to cancel off the external influence. Let us denote the number of  stubborn neighbors of a (non-stubborn) node at site $x$ by $q_x$, and denote the average number of stubborn neighbors among regular nodes by $\bar{q}$.  Let the total number of stubborn nodes be $Q$ and let $\mu$ and $m$ be defined as above, only among ordinary nodes. Also let $\bar{z}$ be the average degree, only taking into account the ordinary nodes \mike{and} not the stubborn ones. 

The equation of motion at (non-stubborn) \mike{node} $x$ becomes
\begin{align}
\dot{c}_x =&~(1-\gamma)B \nonumber \\ 
&+\gamma \left\{ -c_x + \frac{\theta}{z_x+q_x} \left[ - q_x + z_x \left( \frac{2\mu}{\bar{z}}-1 \right) \right] \right. \nonumber\\
&\qquad \left. + \frac{1-\theta}{N+Q}(Nm-Q) \right\}  
.
\label{c_q}
\end{align}
Let us define the constants
\eqq{
\begin{cases}
 \Lambda \stackrel{\text{def}}{=} \frac{1}{N} \sum_x \frac{q_x}{q_x+z_x}  \\
\Omega  \stackrel{\text{def}}{=} \frac{1}{N} \sum_x \frac{z_x q_x}{q_x+z_x} 
\end{cases}
.}{OmegaLambda}
Then using (\ref{c_q}) we get
\begin{align}
\dot{m}= &~-\gamma \left( \frac{\theta+\bar{q}}{1+\bar{q}} \right) m + \left( \frac{2 \theta \gamma \Lambda}{\bar{z}} \right) \mu - \gamma \theta \nonumber\\
& + \left[ B(1-\gamma) - \gamma \bar{q} \frac{1-\theta}{1+\bar{q}} \right]
. \label{mdot_with_B_Q}
\end{align}
For $\mu(t)$ we have:
\begin{align}
\dot{\mu} =&~\left[ \frac{\gamma \bar{z} (1-\theta)}{2(1+\bar{q})}\right]m + \left( -\gamma + \frac{ \gamma \theta \Omega}{\bar{z}} \right) \mu + \frac{\gamma \bar{z}}{2}(1-\theta) \nonumber \\
&+ \frac{\bar{z}}{2} \bigg[(1-\gamma)B -\frac{\gamma (1-\theta) }{(1+\bar{q})}\bar{q} 
\bigg]
.
\label{mudot_with_B_Q}
\end{align}

To cancel the terms involving the external field, it suffices to have
\eqq{
\bar{q}=\frac{B(1-\gamma)}{(1-\theta)\gamma-B(1-\gamma)}
.}{qbar}
This is equivalent to
\eq{
\gamma \bar{q} \frac{1-\theta}{1+\bar{q}} = B(1-\gamma)
,}
so the last terms in~\eqref{mdot_with_B_Q} and~\eqref{mudot_with_B_Q} vanish. This \mike{counterbalances} the external excitation from the dynamics. Combining~\eqref{mdot_with_B_Q} and~\eqref{mudot_with_B_Q}, the equation of motion for the average state becomes
\begin{align}
&\ddot{m} + \dot{m} \gamma \left( \frac{\theta+\bar{q}}{1+\bar{q}} + 1- \frac{ \theta \Omega}{\bar{z}} \right)  \nonumber \\
&+m \gamma^2 
\left[ \frac{\theta + \bar{q}}{1+\bar{q}} \left( 1- \frac{ \theta \Omega}{\bar{z}} \right)- \theta \Lambda \left(\frac{1-\theta}{1+\bar{q}} \right) \right]  \nonumber \\
&= 
\gamma^2 \theta \left( \Lambda(1-\theta)-1+\frac{ \theta \Omega}{\bar{z}}\right)
.
\label{mddot_B_Q}
\end{align}
This is a second order linear differential equation. To analyze the roots, let us define
\eq{
\begin{cases}
\displaystyle K_1 \stackrel{\text{def}}{=} \frac{\theta + \bar{q}}{1+\bar{q}} \medskip \\ 
\displaystyle K_2 \stackrel{\text{def}}{=} 1- \frac{ \theta \Omega}{\bar{z}}  \medskip \\ 
\displaystyle K_3 \stackrel{\text{def}}{=} \theta \Lambda \left(\frac{1-\theta}{1+\bar{q}} \right)
\end{cases}
.}
Then the roots of \eqref{mddot_B_Q} are
\eq{
\begin{cases}
\displaystyle r_1= \frac{-(K_1+K_2)-\sqrt{(K_1-K_2)^2+4 K_3}}{2}  \\ \\
\displaystyle r_2= \frac{-(K_1+K_2)+\sqrt{(K_1-K_2)^2+4 K_3}}{2} .
\end{cases}
}

It is clear that $r_1$ is negative. Now let us examine $r_2$. If $K_1 K_2 \geq K_3$, then 
$r_2$ is negative. \mike{Equivalently}, if the following relationship holds:
\eq{
(\theta + \bar{q}) \bigg(1- \frac{ \theta \Omega}{\bar{z}} \bigg) \geq
\theta \Lambda (1-\theta)
,}
which is equivalent to
\eqq{
\Lambda \leq \left( \frac{1- \frac{ \theta \Omega}{\bar{z}}}{1-\theta}\right) \left(\frac{\theta+\bar{q}}{\theta}\right)
,}{Lambda_ineq}
then $r_2$ will be negative. 
First, note that
\eq{
\Lambda= \frac{1}{N} \sum_x \frac{q_x}{q_x+z_x}  \leq \frac{1}{N} \sum_x 1= 1
.}
\mike{Therefore} $\Lambda$ is less than or equal to unity. Thus, if we show that the right hand side of~\eqref{Lambda_ineq} is at least one, then~\eqref{Lambda_ineq}  holds. To see this, first note that:
 \eq{
\Omega= \frac{1}{N} \sum_x \frac{z_x q_x}{q_x+z_x} 
\leq  \frac{1}{N} \sum_x z_x =\bar{z}
,}  
 or equivalently, 
 \eq{
 \frac{\Omega}{\bar{z}} \leq 1
.}
Now let us re-write the right hand side of~\eqref{Lambda_ineq} \mike{as}
\eq{
\displaystyle \left( \frac{1- \frac{ \theta \Omega}{\bar{z}}}{1-\theta}\right)
\left( \frac{\theta+\bar{q}}{\theta}\right)
.}
For the first factor we have:
\eq{
 \displaystyle  \frac{\Omega}{\bar{z}} \leq 1 \quad \Longrightarrow \quad \theta \frac{\Omega}{\bar{z}} \leq \theta
.}
 Since $\theta \leq 1$, this leads to: 
\eq{
\left( \frac{1- \frac{ \theta \Omega}{\bar{z}}}{1-\theta}\right) \geq 1
.}
For the second factor, we have: 
\eq{
\left( \frac{\theta+\bar{q}}{\theta}\right) = 1+\frac{\bar{q}}{\theta} \geq 1
.}
So we have found that 
\eq{
\Lambda \leq 1 \leq  \left( \frac{1- \frac{ \theta \Omega}{\bar{z}}}{1-\theta}\right) \left(\frac{\theta+\bar{q}}{\theta}\right)
,}
which means that~\eqref{Lambda_ineq} holds. Thus $r_2 \leq 0$. 

As we have shown, both roots of~\eqref{mddot_B_Q} are negative, so the homogeneous answer decays, and the particular response prevails. We obtain an expression for the particular response by setting the time derivatives in~\eqref{mddot_B_Q} equal to zero. This gives the steady-state solution:
\eqq{
m(\infty)= \frac{\theta \left( \Lambda(1-\theta)-1+\frac{ \theta \Omega}{\bar{z}}\right)}
{ 
\left[ \frac{\theta + \bar{q}}{1+\bar{q}} \left( 1- \frac{ \theta \Omega}{\bar{z}} \right)- \theta \Lambda \left(\frac{1-\theta}{1+\bar{q}} \right) \right]}
.}{mpred}

The denominator is positive as we showed above. For the numerator, note that the following holds: 
\eq{
\Lambda \leq 1 \leq  \left( \frac{1- \frac{ \theta \Omega}{\bar{z}}}{1-\theta}\right) 
,}
or equivalently, 
\eq{
\Lambda (1-\theta) \leq  \left( 1- \frac{ \theta \Omega}{\bar{z}}\right) 
}
\mike{Thus,} the numerator of~\eqref{mpred} is negative, \mike{and} $m(\infty)$ is negative. Remember that the exogenous stimulus was trying to bias the states towads $s=+1$ and the anti-conformists were trying to pull the states towards $s=-1$. Equation~\eqref{mpred} states that the average steady-state opinion of the whole population is negative, which is in favor of the stubborn nodes \mike{as long as~\eqref{qbar} holds}.

\begin{figure}[ht]
  \centering
  \includegraphics[width=90mm, height=45mm]{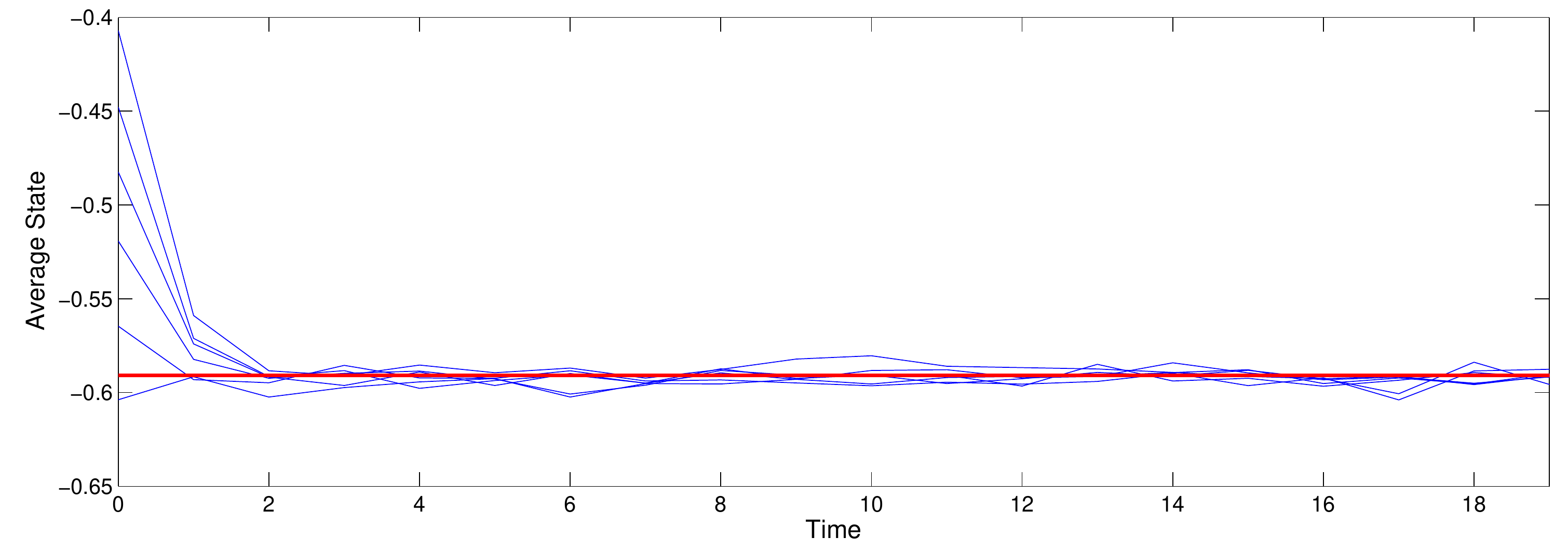}
  \caption[Figure 2]%
  {m(t) with respect to time, for different values of $\rho$ between zero and $0.9$ (uniform), the red line is the prediction of (\ref{mpred}). The underlying graph is random with 500 nodes. The results are averaged over 100 Monte Carlo trials.}
\label{figmbq}
\end{figure}

We conclude that the distribution of stubborn nodes controls the equilibrium bias through $\Lambda$ and $\Omega$, given that (\ref{qbar}) holds. Figure~\ref{figmbq} shows the simulation results and convergence to what \eqref{mpred} predicts.

\mike{Now let us examine a special case where the number of stubborn neighbors of a non-stubborn node is a constant fraction of its degree. In particular, let us assume that there is a constant $\alpha \in (0,1)$ such that $q_k$, the number of stubborn neighbors of a non-stubborn node with degree $k$, is equal to $q_k = \alpha k$ for all $k$. Then we have} 
\eqq{ 
\alpha=\frac{\frac{B}{\bar{z}}(1-\gamma)}{(1-\theta)\gamma-B(1-\gamma)}
.}{alpha_final}
For the parameters in (\ref{OmegaLambda}) we have: 
\eq{
\begin{cases}
 \Lambda =  \frac{\alpha}{\alpha+1}  \\
\Omega  = \left(\frac{\alpha}{\alpha+1}\right)\bar{z}
\end{cases}
.}
The final  average state becomes: 
\eq{
m(\infty)= \frac{\frac{- \theta}{\alpha+1}}
{ 
\frac{\theta + \alpha \bar{z} + \alpha^2 \bar{z} (1-\theta)}{(1+\alpha \bar{z})(1+\alpha)}
} < 0
,}
which is negative, meaning that the bias leans towards the stubborn nodes.   Finally, note that~\eqref{alpha_final} implies that to have~${\alpha<1}$, we must have
\eq{
\displaystyle \gamma \geq \frac{B(1+\frac{1}{\bar{z}})}{(1-\theta)+B(1+\frac{1}{\bar{z}})},
}
which means that the population must care less than some threshold for the inflicted external bias.\\

\section{Summary and Discussion} \label{sec:conclusion}

We considered the problem of binary states, or opinions, among nodes that are connected through a network of arbitrary degree distribution. We modelled the social influence by two distinct parts. Each node finds the fraction of nodes who are adjacent to it and disagree with it, and also the fraction of the nodes in the whole network that disagree with it. It then takes a convex combination of those fractions, to find a probability,  and at each timestep, flips its opinion with this probability. We solved for the probability of consensus on either state, and the expected time to reach consensus, as a function of initial conditions. We also found the time evolution of the average state over the whole population.

Then we added an external influence to the model, that tries to bias all nodes towards \mike{the state $+1$}. We solved for the average state of the system in time. 

To \mike{counterbalance} the effect of the exogenous bias, we introduced stubborn nodes. These nodes constantly oppose the external bias. We found the equilibrium average state of the system, and the necessary conditions so that the stubborn nodes can succeed in defeating the external field. 

Possible extensions of the model include using a network topology with given degree correlations, and using it to improve the mean field assumption. Also, it will be more realistic, albeit analytically somewhat formidable, if the influence between each pair of nodes \mike{is} a function of the distance between them. Alternatively, each node, when accounting for the state of other nodes, can assign different weights to them (see~\cite{voter_weighted}), that can be functions of other types of centralities. In this case, the more central a node is, the more influence it exerts on the nodes it interacts with. Upon existence of the external field, nodes can be given different levels of exposure, or different degrees of resistance against it. \mike{This means that stubbornness can be a quality that exists in all nodes but to varying degrees.} 

Another useful modification, especially for marketing purposes, is as follows. Suppose $s=-1$ \mike{signifies} that the node \mike{has} not bought some product, or \mike{has} not watched some movie. Then, one can modify the model so that nodes \mike{can not revert} back once they \mike{have} adopted $s=+1$ (\mike{the class of} so-called ``susceptible-infected'' models in the epidemiology literature). This means that, for example, once \mike{a node has seen a} movie, \mike{it} cannot un-see it. This will render the model readily testable against empirical data.

\bibliographystyle{IEEEtran}
\bibliography{bib2}

\end{document}